\documentclass[letterpaper, 10 pt, conference]{ieeeconf} 
\usepackage{amssymb, amsmath, latexsym, amsfonts, mathrsfs, amsbsy,mathtools,relsize}
\usepackage{float}
\usepackage{algorithm}
\usepackage{algorithmic}
\usepackage[english]{babel}
\usepackage{soul}
\usepackage[font=small,labelfont=bf]{caption}
\usepackage{graphicx}
\usepackage[utf8]{inputenc}
\usepackage{epsfig, color}
\usepackage[sort]{cite}
\usepackage{listings}
\usepackage{subfloat}
\usepackage{setspace}
\usepackage{graphicx}
\usepackage{makeidx, glossaries}
\usepackage{url}
\usepackage{pgfplots}
\usepackage{flushend}
\usepackage{adjustbox}
\usepackage[caption=false, font=footnotesize]{subfig}
\usepackage{stmaryrd}
\usepackage{enumerate}
\usepackage{epigraph}
\usepackage{multirow}
\usepackage{footnote}
\usepackage{bbm}

\usepackage[T1]{fontenc}
\usepackage{calligra}

\usepackage{algorithm, algorithmic}
\usepackage{lipsum}

\usepackage{physics}
\usepackage{color}
\usepackage{array,booktabs,calc}
\usepackage{float}
\usepackage{graphics} 
\usepackage{epsfig} 
\usepackage{multirow}
\usepackage{amsmath}
\usepackage{setspace}
\usepackage{amsbsy}
\usepackage{supertabular}
\usepackage{tkz-graph}
\usepackage{etex}
\usepackage{pgf, tikz}
\usepackage{bm}
\usetikzlibrary{shapes,shapes.geometric,arrows,fit,calc,positioning,automata,through,intersections}
\tikzset{diamond state/.style={draw,diamond}}

\usepackage{capt-of}
\usepackage{calligra}

\usepackage{setspace}

\usepackage{subfiles}

\usepackage{pgf, tikz}

\usepackage{amsmath,bm,amsfonts}
\usepackage{pgfplots}
\pgfplotsset{compat=1.18}
\usetikzlibrary{shapes.geometric, arrows.meta, positioning, calc, shadows}

\definecolor{boxblue}{RGB}{240, 245, 250}
\definecolor{headerblue}{RGB}{50, 100, 150}
\definecolor{accentblue}{RGB}{0, 114, 189}
\definecolor{alarmred}{RGB}{217, 83, 25}

\usetikzlibrary{shapes,shapes.geometric,arrows,fit,calc,positioning,automata,through,intersections}
\tikzset{diamond state/.style={draw,diamond}}
\usetikzlibrary{positioning,calc}
\usepackage{smartdiagram}

\usepackage{graphicx}

\newenvironment{list4}{
	\begin{list}{$\bullet$}{%
			\setlength{\itemsep}{0.05cm}
			\setlength{\labelsep}{0.2cm}
			\setlength{\labelwidth}{0.3cm}
			\setlength{\parsep}{0in} 
			\setlength{\parskip}{0in}
			\setlength{\topsep}{0in} 
			\setlength{\partopsep}{0in}
			\setlength{\leftmargin}{0.16in}}}
	{\end{list}}

\usepackage{tikz}
\usetikzlibrary{arrows.meta,calc,positioning,decorations.pathreplacing}

\IEEEoverridecommandlockouts                              

\title{\LARGE \bf
Model-Based Detection of Anomalous Events in Submarine Cables Using
Distributed Deformation Sensing and Kalman Filtering
}

\author{Camilla Fioravanti, Bianca Mazz\'a, Marta Menci, Gabriele Oliva$^{*}$, and Roberto Setola
\thanks{Department of Engineering, Universit\`a Campus Bio-Medico di Roma, via \'Alvaro del Portillo 21, 00128, Rome, Italy. }
\thanks{$^*$ corresponding author. Email: g.oliva@unicampus.it}
\thanks{This work was partly supported by project VIGIMARE, funded by the European Union under grant no. 101168016. Views and opinions expressed are however those of the authors only and do not necessarily reflect those of the European Union. Neither the European Union nor the granting authority can be held responsible for them. This work was partly supported by Italian National project IMPROVE, funded by the Italian Ministry of Defense under grant no. 20711.}
}

\begin{document}

\maketitle
\thispagestyle{empty}
\pagestyle{empty}

\begin{abstract}
Submarine power and telecommunication cables constitute critical global infrastructure, yet they remain vulnerable to mechanical damage caused by maritime activities and intentional tampering. Continuous monitoring of these assets is therefore essential for early detection of anomalous events. This paper proposes a model-based framework for real-time anomaly detection in submarine cables using spatially distributed deformation measurements along the cable. 
The cable is modeled as a tensioned structure governed by a damped wave equation with fixed boundary conditions. A finite-dimensional state-space representation is obtained through spatial discretization, enabling the use of a Kalman filter to estimate the cable's dynamic state under stochastic environmental disturbances. Anomaly detection is then formulated as a statistical hypothesis test applied to the innovation sequence of the filter. Numerical simulations indicate that the proposed framework can reliably identify localized disturbances while remaining robust to ambient environmental excitation.
\end{abstract}

\section{Introduction}
Submarine power and communication cables are critical assets for
modern society, as they support digital connectivity, interconnection
of energy systems, and the integration of offshore infrastructures
~\cite{carter2009submarine,kavanagh2025achieving}.
At the same time, they are exposed to a wide spectrum of hazards,
including anchor drags, fishing activity, seabed instability, fatigue,
and potentially intentional tampering
~\cite{osthoff2017impact,guilfoyle2022final,yoon2013safety}.
These factors make continuous monitoring of cable integrity a key
requirement for both operational reliability and infrastructure
protection.

Recent works have investigated monitoring strategies for subsea
cable systems and marine infrastructures, where long flexible lines
operate under complex environmental loading conditions~\cite{yang2024analytical}.
A key challenge is that anomalous mechanical interactions are often
masked by environmental excitation generated by waves, currents,
and seabed contact~\cite{irvine1974linear}.
Effective monitoring approaches should therefore rely on measurements
that directly capture the cable mechanical response over extended
spatial domains.
In particular, distributed optical fiber sensing technologies enable
dense measurements of strain, deformation, or vibration along very long
structures, often with meter-scale spatial sampling over kilometer-scale
distances
\cite{hartog2017introduction,lu2019distributed,lindsey2021fiber}.
These technologies have been widely investigated for structural health
monitoring applications
\cite{murayama2013structural}, and have proven particularly suitable
for monitoring cable-like structures, since they allow spatially
continuous strain measurements over long distances and enable the
detection of localized structural anomalies along the cable span
\cite{masoudi2018dynamic}.
In the specific case of submarine and subsea infrastructures,
distributed acoustic sensing (DAS) has already shown the potential to
detect vessel activity, anchoring, trawling, seismic events, and other
external disturbances by turning the fiber itself into a continuous
sensing array
\cite{waagaard2022experience,lindsey2021fiber}.
Recent works further highlight how distributed sensing technologies can
support advanced monitoring and reliability assessment frameworks for
cable structures and other elongated infrastructures
\cite{li2025research}.

Despite these sensing advances, interpreting distributed measurements
remains challenging.
Purely data-driven approaches can be effective when large amounts of
labeled data are available, but in safety-critical monitoring they may
suffer from limited interpretability and poor generalization beyond the
conditions observed during training.
Recent reviews in structural health monitoring therefore highlight the
need to combine data with physical knowledge rather than relying
exclusively on black-box inference~\cite{cha2024deep}.
This issue is particularly critical for submarine cables, where
anomalous events must be distinguished from strong environmental
variability under scarce fault data.
For this reason, a physics-based monitoring strategy is particularly
appealing.
Classical cable dynamics literature shows that, under small transverse
displacements and approximately constant tension, cable motion can be
described through distributed-parameter models based on wave-type
equations \cite{irvine1974linear}.
Related works have also investigated control and estimation strategies
based on hyperbolic PDE models of cable dynamics, highlighting the
importance of physically grounded distributed-parameter descriptions
for long flexible structures \cite{baudouin2018robust}.
Such models retain the essential propagation, reflection, and damping
mechanisms that govern the response of slender tensioned structures,
while remaining sufficiently structured to support estimation and
detection algorithms.
In this sense, distributed-parameter modeling provides a natural bridge
between spatially distributed sensing and real-time statistical
monitoring.

\noindent {\bf Contribution.}
This paper proposes a model-based framework for the real-time detection
of anomalous mechanical events in submarine cables from distributed
deformation measurements. The cable is modeled as a tensioned structure
governed by a damped wave equation with fixed boundary conditions, and a
finite-dimensional linear state-space representation is derived through
spatial discretization. On top of this model, a Kalman filter is used to
estimate the cable state under stochastic environmental excitation, while
anomaly detection is formulated as a statistical test on the innovation
sequence. The main contribution of the paper is the integration of a
physically grounded distributed-parameter cable model with
innovation-based statistical monitoring, yielding a computationally
tractable framework for detecting localized disturbances in the presence
of environmental variability. The scope of the paper is deliberately focused on detection, while event classification, extensive comparison with alternative detectors, and field-data validation are left for future extensions.

\color{black}
\section{Cable--Sensor System Physical Modeling}

Submarine communication and power cables deployed on the seabed
are continuously exposed to environmental excitation and
potential external interactions.
Modern sensing technologies enable spatially distributed
measurements along long cable structures, allowing mechanical
perturbations to be monitored through the dynamic response of
the cable itself.
In this section, we introduce a physically grounded model
for the coupled cable and measurement system.

\noindent {\bf Cable Dynamics: Distributed Parameter Model.}
Classical analyses of suspended cable vibrations show that, under small transverse displacements, the dynamics of a tensioned cable can be approximated by a tension-dominated wave equation derived from distributed force balance along the cable \cite{irvine1974linear}.
 Accordingly, \color{black} we model the submarine cable as a slender, tensioned structure
undergoing small transverse displacements. 
Let $u(x,t)$ denote the transverse displacement
at spatial coordinate $x \in [0,L]$ and time $t \ge 0$.

Assuming small deformations, approximately constant axial tension,
and linear viscous damping due to hydrodynamic effects--assumptions
commonly adopted in first-order models of cable vibrations under
small transverse perturbations \cite{irvine1974linear}--the cable dynamics are described by the damped wave equation with damping and external source terms, i.e.,
\begin{equation}
\rho \frac{\partial^2 u}{\partial t^2}(x,t)
=
T \frac{\partial^2 u}{\partial x^2}(x,t)
-
c \frac{\partial u}{\partial t}(x,t)
+
f(x,t),
\label{eq:damped_wave_model}
\end{equation}
where: $\rho$ denotes the effective linear mass density, $T$ is the axial tension, $c$ is a viscous damping coefficient accounting for hydrodynamic drag, and $f(x,t)$ represents external distributed forcing.

The validity of this linear model is therefore limited to operating regimes in which transverse cable excursions are small relative to the monitored span and do not significantly change the axial tension or the cable--seabed contact conditions. Large deformations would require nonlinear cable dynamics, which fall beyond the scope of this paper. 

The term $f(x,t)$ models environmental excitation
(e.g., currents or seabed interactions)
as well as possible localized disturbances due to
external mechanical interactions.
In particular, we consider an environmental excitation term
\begin{equation}
f(x,t)
=
\eta(x,t)
+
\sum_{p=1}^{P}
\delta(x - x_p) a_p(t),
\end{equation}

where $\eta(x,t)$ is a stochastic distributed excitation,
while $a_p(t)$ represents localized perturbations
at spatial positions $x_p$, $p\in\{1,\ldots P\}$ and $P$ denotes the number of discrete locations where localized external disturbances are applied. Here $\delta(\cdot)$ denotes the Dirac delta distribution, used as an idealized point-load model for a force concentrated at $x_p$; in the discretized model, this term is implemented by assigning the corresponding load to the closest grid node or to a small group of neighboring nodes.
In order to adopt the above model, appropriate boundary conditions must be specified.
In this work, we model a finite, tensioned cable segment securely 
anchored at both extremities. 
This corresponds to fixed (Dirichlet) 
boundary conditions. Without loss of generality we here consider null Dirichlet boundary conditions, i.e., $u(0,t) = u(L,t) = 0$.

\noindent {\bf Distributed Deformation Measurement Model.}
Let us now characterize the measurement model. In particular, we consider spatially distributed sensing technologies, which enable measurements
of deformation or strain along long cable structures \cite{hartog2017introduction}.
Such measurements provide information about the dynamic response
of the cable and can be used to detect localized disturbances
along its length.
Specifically, under small-deformation assumptions, we can approximate local measurements as proportional to the spatial gradient of the
cable displacement field.
Thus, an idealized continuous measurement
at position $x$ is proportional to \mbox{$\varepsilon(x,t)
\approx
\frac{\partial u}{\partial x}(x,t),$} where $\varepsilon(x,t)$ denotes the axial strain of the cable, i.e., the spatial derivative of the transverse displacement field $u(x,t)$.

In practice, distributed sensing systems measure strain over a finite
\emph{gauge length} $g$, so that the reported signal represents a spatially
averaged strain over an interval of length $g$.
The measurement at a sensing channel centered at $x_j$ can therefore
be modeled as
\begin{equation}
y_j(t)
=
\kappa
\frac{1}{g}
\int_{x_j - g/2}^{x_j + g/2}
\frac{\partial u}{\partial x}(\xi,t)
\, d\xi
+
\omega_j(t),
\label{eq:continuous_measurement}
\end{equation}
where $\kappa$ is a calibration constant relating deformation to the
measured signal, and
$\omega_j(t)$ represents measurement noise.
By exploiting the fundamental theorem of calculus,
Eq.~\eqref{eq:continuous_measurement} can be rewritten as
\begin{equation}
y_j(t)
=
\kappa
\frac{
u(x_j + g/2,t)
-
u(x_j - g/2,t)
}{g}
+
\omega_j(t),
\end{equation}
which highlights the finite-difference nature of the distributed measurement.
The gauge length $g$ acts as a spatial averaging window. A smaller $g$ improves the spatial resolution of the measurement and helps detect highly localized disturbances, but it may also increase sensitivity to sensor noise and local installation effects. Conversely, a larger $g$ improves averaging but may attenuate disturbances whose spatial footprint is shorter than, or comparable to, the gauge length. Thus, in practice, $g$ should be selected consistently with the smallest anomaly length scale that the monitoring system is expected to detect.

By stacking $m$ sensing channels we obtain the vector measurement model \mbox{$y(t)
=
\mathcal{C}[u(\cdot,t)]
+
\omega(t),$} where $\mathcal{C}$ is a linear operator combining
spatial differentiation and gauge-length averaging.
Therefore, we can observe that the measurement equation is linear in the displacement field.
Together with the distributed cable dynamics,
this yields a linear infinite-dimensional
state-output system, which will be approximated
by a finite-dimensional linear model in the next section. 
The placement of the sensing channels affects observability and localization: uniformly distributed channels are a natural baseline for long homogeneous spans, whereas denser placement should be used near expected high-risk regions such as anchoring areas, landfalls, joints, or zones with complex seabed interaction.
\section{Finite-Dimensional Approximation}

In order to enable state estimation and statistical monitoring,
the distributed-parameter model introduced in the previous section
is approximated by a finite-dimensional linear system.
This is achieved by discretizing the spatial domain
via finite-difference techniques \cite{leveque2007finite} and subsequently rewriting
the dynamics in state-space form using a time-discretization scheme whose step size is selected according to the stability conditions reported below.

\noindent {\bf Spatial Discretization via Finite Differences.}
Let us consider the spatial interval $[0,L]$. We consider a cable segment that is firmly anchored at its extremities; therefore, we use the aforementioned Dirichlet null boundary conditions.

We can introduce a uniform grid of $N$ equispaced interior points 
$x_i = i\Delta x$ for $i = 1,\dots,N$, where the spatial step is defined 
relative to the fixed boundaries as $\Delta x = \frac{L}{N+1}$. Then, let the 
displacement at the interior nodes be approximated by the vector ${\bm u}(t) = [u_1(t),\dots,u_N(t)]^\top$, where $u_i(t) \approx u(x_i,t)$.
The grid size should be chosen fine enough to resolve the shortest wavelength and spatial anomaly scale of interest, as well as the effective gauge-length averaging introduced by the sensor. Increasing $N$ improves the spatial localization and the readability of space--time diagrams, but it also increases the dimension of the filtering problem.
The second-order spatial derivative is approximated using centered finite differences. 
Because the boundary nodes are strictly zero ($u_0 = u_{N+1} = 0$), the wave 
reflections are naturally captured by the symmetric, tridiagonal 
second-difference matrix $D_2 \in \mathbb{R}^{N\times N}$
\begin{equation}
(D_2)_{ij} = \begin{cases}
\frac{1}{\Delta x^2}, & \mbox{if } |i-j|=1 \\
-\frac{2}{\Delta x^2}, & \mbox{if } i=j \\
0 & \mbox{otherwise}.
\end{cases}
\end{equation}
By substituting this spatial approximation into the damped wave equation, 
we obtain a finite-dimensional system of coupled ordinary differential equations
\begin{equation}
\rho \ddot{{\bm u}}(t)
=
T D_2 {\bm u}(t)
-
c \dot{{\bm u}}(t)
+
{\bm w}(t),
\label{eq:semi_discrete}
\end{equation}
where ${\bm w}(t) \in \mathbb{R}^N$ represents distributed forcing and process disturbances, and $T$ and $c$ are defined in Eq.~\eqref{eq:damped_wave_model}. 

The distributed measurement model introduced in the previous section
can also be expressed in terms of the discretized displacement vector
${\bm u}(t)$. In particular, the measurements correspond to spatial
finite-difference operations applied to the displacement field,
resulting in a linear observation model of the form \mbox{${\bm y}(t) = C_u {\bm u}(t) + {\bm \omega}(t)$}, where $C_u \in \mathbb{R}^{m \times N}$ represents the discrete gauge-length
difference operator and ${\bm \omega}(t) \in \mathbb{R}^{m}$ denotes
measurement noise. 
\noindent {\bf Continuous-Time State-Space Representation.}
To express the dynamics in finite-dimensional state-space form, let us first define the velocity vector \mbox{${\bm v}(t) = \dot{{\bm u}}(t),$} and introduce the extended state vector \mbox{${\bm x}(t)
=
\begin{bmatrix}
{\bm u}^\top(t)&
{\bm v}^\top(t)
\end{bmatrix}^\top
\in \mathbb{R}^{2N}.$}
Thus, Eq.~\eqref{eq:semi_discrete} can be rewritten in first-order, 
continuous-time state-space form as
\begin{equation}
\dot{{\bm x}}(t)
=
A {\bm x}(t)
+
B {\bm w}(t),
\label{eq:ct_statespace}
\end{equation}
where the system matrices are defined as
\begin{equation}
A =
\begin{bmatrix}
0 & I \\
\frac{T}{\rho} D_2 & -\frac{c}{\rho} I
\end{bmatrix}, \quad 
B =
\begin{bmatrix}
0 \\
\frac{1}{\rho} I
\end{bmatrix}.
\end{equation}

Notice that the resulting model is linear and time-invariant and that the system matrix $A$ encapsulates both the tension-driven wave propagation
(through the discrete Laplacian $D_2$) and the hydrodynamic damping ($-(c/\rho)I$).
Since the measurements depend only on the cable deformation and not on its velocity, the observation equation can be written in terms of the
state vector as \mbox{${\bm y}(t) = C {\bm x}(t) + {\bm \omega}(t)$}, with \mbox{$C = [\, C_u \;\; 0_{m\times N} \,]\in \mathbb{R}^{m \times 2N}$}.
This completes the continuous-time state-space representation of the
discretized cable dynamics together with the distributed measurement model.

\noindent {\bf Discrete-Time Model.}
For digital implementation and state estimation, Eq.~\eqref{eq:ct_statespace} 
is discretized in time with a sampling period $\Delta t$. In particular, we exploit the 
Forward Euler integration scheme to obtain a discrete-time 
linear representation. The state transition matrix $A_d$ and 
the input mapping matrix $B_d$ are defined as $A_d = I + \Delta t A$ and $B_d = \Delta t B.$
The resulting discrete-time system is described by the following state-space equation
\begin{equation}
\begin{cases}
  {\bm x}[k+1] &= A_d {\bm x}[k] + B_d {\bm w}[k],  \\
  {\bm y}[k] &= C {\bm x}[k] + {\bm \omega}[k],
\end{cases}
\label{eq:discrete_statespace}
\end{equation}
where ${\bm w}[k]$ and ${\bm \omega}[k]$ represent the discrete-time process and measurement noises, respectively. This yields a linear model suitable for Kalman filtering and innovation-based anomaly detection.

In order to ensure stability of the proposed numerical scheme, two conditions arise.
Due to the hyperbolic nature of equation \eqref{eq:damped_wave_model}, the Courant-Friedrichs-Lewy (CFL) condition requires \mbox{$\Delta t \le \frac{\Delta x}{\sqrt{\frac{T}{\rho}}}.$} 
Moreover, the choice of Forward Euler time integration induces an additional restriction due to the stiff damping term. Specifically, stability is ensured by choosing \mbox{$\Delta t \le \frac{2\rho}{c}.$}

\section{State Estimation and Anomaly Detection}
In the previous section, we derived the finite-dimensional discrete-time model. Since it is linear, we can exploit the Kalman filter~\cite{kalman1960new} to perform the state estimation that will be the basis for the anomaly detection procedure.

Let us first consider the discrete-time system in Eq.~\eqref{eq:discrete_statespace}, and assume ${\bm w}[k]$ and ${\bm \omega}[k]$ are zero-mean,
independent Gaussian noise sequences with covariances
$Q_d$ and $R$, respectively.
These Gaussianity and independence assumptions are used as a design model for deriving the Kalman filter and the nominal test threshold. In practice, $Q_d$ and $R$ can be selected from sensor specifications and nominal-data calibration, and then refined by innovation covariance matching. If the environmental forcing is non-Gaussian or time-correlated, the same innovation signal can still be monitored, but the threshold should be calibrated empirically or after whitening/robustification rather than relying only on the nominal chi-square approximation.
Let $\hat{{\bm x}}_{k|k-1}$ denote the one-step state prediction and
$P_{k|k-1}$ the associated covariance matrix.
The Kalman filter prediction equations are
\begin{equation}
\hat{{\bm x}}_{k|k-1}
=
A_d \hat{{\bm x}}_{k-1|k-1},
\end{equation}
\begin{equation}
P_{k|k-1}
=
A_d P_{k-1|k-1} A_d^\top + B_d Q_d B_d^\top.
\label{eq:Ppred}
\end{equation}
The innovation equation is
\begin{equation}
{\bm e}[k]
=
{\bm y}[k] - C \hat{{\bm x}}_{k|k-1}.
\end{equation}
and the update equations are
\begin{equation}
S[k]
=
C P_{k|k-1} C^\top + R,
\end{equation}
\begin{equation}
K[k]
=
P_{k|k-1} C^\top S^{-1}[k],
\end{equation}
\begin{equation}
\hat{{\bm x}}_{k|k}
=
\hat{{\bm x}}_{k|k-1} + K[k] {\bm e}[k],
\end{equation}
\begin{equation}
P_{k|k}
=
(I - K[k] C) P_{k|k-1}.
\end{equation}

Under nominal operating conditions, i.e., when the system model is correct, the disturbances are captured by ${\bm w}[k]$, and the measurement noise is accurately modeled by ${\bm \omega}[k]$, the innovation sequence ${\bm e}[k]$ satisfies
$\mathbb{E}[{\bm e}[k]] = 0,$
and has covariance $S[k]$.
More precisely, nominal operation here means the absence of persistent localized external forces, approximately constant model parameters over the analysis window, and stationary zero-mean environmental and measurement disturbances. These assumptions are idealizations, but they are realistic for short monitoring windows in which the dominant environmental variability is captured by the process-noise model.
Therefore, the innovation represents the portion of the measurement that cannot be predicted from past data, given the assumed model.
On the other hand, unexpected external interactions, unmodeled disturbances,
or parameter variations introduce a systematic bias in the innovation sequence.
In particular, the presence of a persistent disturbance generally leads to
$\mathbb{E}[{\bm e}[k]] \neq 0$.

To detect such deviations, we consider the sample mean of the innovation over a sliding window of length $M$, i.e., \mbox{$\bar{{\bm e}}[k]
=
\frac{1}{M}
\sum_{i=k-M+1}^{k}
{\bm e}[i],$}
and we assume that, under nominal conditions,
$\bar{{\bm e}}[k]$ converges to zero as $M$ increases. According to this, any statistically significant deviation of $\bar{{\bm e}}[k]$
from zero indicates a possible model mismatch
or anomalous mechanical interaction with the cable. Thus, a zero-mean statistical test provides a simple and
computationally efficient mechanism for
real-time anomaly detection. 
Specifically, we test the innovation mean (e.g., see~\cite{johnson2002applied}) under the following hypotheses
\begin{align}
\mathcal{H}_0 &: \quad \mathbb{E}[{\bm e}[k]] = 0
\quad \text{(nominal operation)}, \\
\mathcal{H}_1 &: \quad \mathbb{E}[{\bm e}[k]] \neq 0
\quad \text{(anomalous condition)}.
\end{align}
The innovation mean is estimated through the sample mean $\bar{{\bm e}}[k]$.
Notice that, under $\mathcal{H}_0$,
the innovation is zero-mean with covariance $S[k]$.
Assuming weak temporal correlation,
the sample mean has approximate covariance \mbox{$\mathrm{Cov}(\bar{{\bm e}}[k])
\approx
\frac{1}{M} S[k].$}
A scalar test statistic can then be defined as \mbox{$T_k
=
M\,\bar{{\bm e}}^\top[k] S^{-1}[k] \bar{{\bm e}}[k], $}
and under $\mathcal{H}_0$ and Gaussian assumptions,
$T_k$ approximately follows a chi-square distribution
with $m$ degrees of freedom,
where $m$ is the measurement dimension~\cite{johnson2002applied}.
According to this scheme, an alarm is raised whenever \mbox{$T_k > \gamma,$}
where $\gamma$ is selected as the $(1-\alpha)$ quantile of the
$\chi^2_m$ distribution in order to achieve a desired false-alarm
probability $\alpha$.
Therefore, the proposed detector identifies persistent deviations
from the nominal model, corresponding to sustained
external interactions or structural perturbations
affecting the cable.

The dominant per-sample computational cost is due to the
Kalman-filter covariance recursion. Let $n=2N$ be the state dimension
and $m$ the number of sensing channels. In a dense implementation,
the covariance prediction $P_{k|k-1}$, computed via Eq.~\eqref{eq:Ppred}, requires products between $n\times n$ matrices, and therefore scales as
$\mathcal{O}(n^3)$. The innovation covariance $S[k]$ requires multiplying $C\in\mathbb{R}^{m\times n}$ by
$P_{k|k-1}\in\mathbb{R}^{n\times n}$, with cost
$\mathcal{O}(n^2m)$, followed by multiplication by $C^\top$, with cost
$\mathcal{O}(nm^2)$. Finally, the Kalman gain $K[k]$
requires the product $P_{k|k-1}C^\top$, with cost
$\mathcal{O}(n^2m)$, and the inversion or factorization of
$S[k]\in\mathbb{R}^{m\times m}$, with cost $\mathcal{O}(m^3)$.
Thus, neglecting the lower-order term $\mathcal{O}(nm^2)$ when
$m\le n$, the dense per-sample complexity is
$\mathcal{O}(n^3+n^2m+m^3)$, with memory requirement
$\mathcal{O}(n^2)$. The sliding-window innovation mean and the
corresponding quadratic test are cheaper, scaling as
$\mathcal{O}(m)$ and $\mathcal{O}(m^2)$, respectively.
\color{black}
\section{Numerical Simulation Study}
\label{sec:simulations}

This section illustrates the proposed modeling and detection framework through numerical simulations based on realistic cable parameters. The simulation illustrates both the physical wave propagation and the statistical anomaly detection capabilities of the Kalman filter.

\noindent {\bf Simulation Setup.}
For our simulation setup, we consider four main aspects. 
\begin{list4}
    \item {Cable parameters:}
we consider a submarine cable segment of length $L = 1000\,\text{m}$. The effective linear mass density is set to $\rho = 15\,\text{kg/m}$, while the axial tension is chosen as $T = 40000\,\text{N}$. A viscous damping coefficient $c = 800\,\text{N\,s/m}$ is introduced to model hydrodynamic drag along the seabed.

\item{Noise modeling and sensor configuration:}
process noise, representing oceanic environmental forcing, is modeled as a zero-mean Gaussian excitation acting on the state with covariance $Q = q I$, where $q = 20$. Measurement noise is assumed Gaussian with covariance $R = r I$, where $r = 0.05$.

\item{Spatial discretization:}
the spatial domain is discretized using $N = 40$ equispaced interior grid points, resulting in a spatial resolution of $\Delta x = L / (N+1) \approx 24.39\,\text{m}$. 
We impose strict fixed boundary conditions (Dirichlet conditions, where $u(0,t) = 0$ and $u(L,t) = 0$) to accurately model a finite, tensioned cable segment where wave reflections occur at the endpoints. 
This value of $N$ is sufficient for the illustrative disturbance considered here, whose spatial footprint is broad compared with $\Delta x$; a finer discretization would improve localization and visualization at the expense of the filtering cost discussed above.

\item{Temporal discretization:}
the sampling period is set to $\Delta t = 0.02\,\text{s}$, satisfying the CFL and Euler stability conditions discussed above. The continuous-time 
state-space system is mapped to discrete time using the Forward Euler 
approximation, ensuring a direct correspondence between the theoretical 
framework established in Section III and the numerical implementation.
\end{list4}

We assume that a total of $m = 10$ sensing channels are uniformly distributed across the active interior nodes of the cable. Each channel provides a distributed deformation measurement corresponding to a finite-difference approximation of the spatial gradient of the displacement field. Uniform placement is used as a neutral baseline; in practical deployments, channel spacing should be selected so that the expected anomaly footprint is covered by multiple measurements, while high-risk sections can be monitored with denser effective sampling.
%
Under nominal conditions, the cable is excited solely by stochastic environmental forcing. The Kalman filter is initialized with $P_0 = I$.


To simulate an anomalous event, a localized external force is injected at the physical center of the cable ($x_p = 500\,\text{m}$) operating between $t = 5\,\text{s}$ and $t = 10\,\text{s}$. This step-function disturbance models a sustained, aggressive mechanical interaction, such as an anchor strike or deliberate physical tampering.
In this scenario, anomaly detection is performed via a sliding-window mean test on the 
innovation sequence. Under the null hypothesis $\mathcal{H}_0$, the test 
statistic $T_k$ follows a chi-square ($\chi^2$) distribution with $m=10$ 
degrees of freedom. To achieve a high confidence level and 
suppress false alarms from random ocean noise spikes, the detection 
threshold is set to $\gamma = 25$. This value corresponds to a significance 
level of approximately $\alpha = 0.005$, implying that the probability 
of a false alarm occurring due to stochastic environmental noise alone 
is limited to $0.5\%$. This choice ensures that the detector 
remains robust against the ambient noise floor while maintaining 
high sensitivity to the systematic biases introduced by physical tampering.

\begin{figure}[htpb]
    \centering
    \includegraphics[width=0.98\columnwidth]{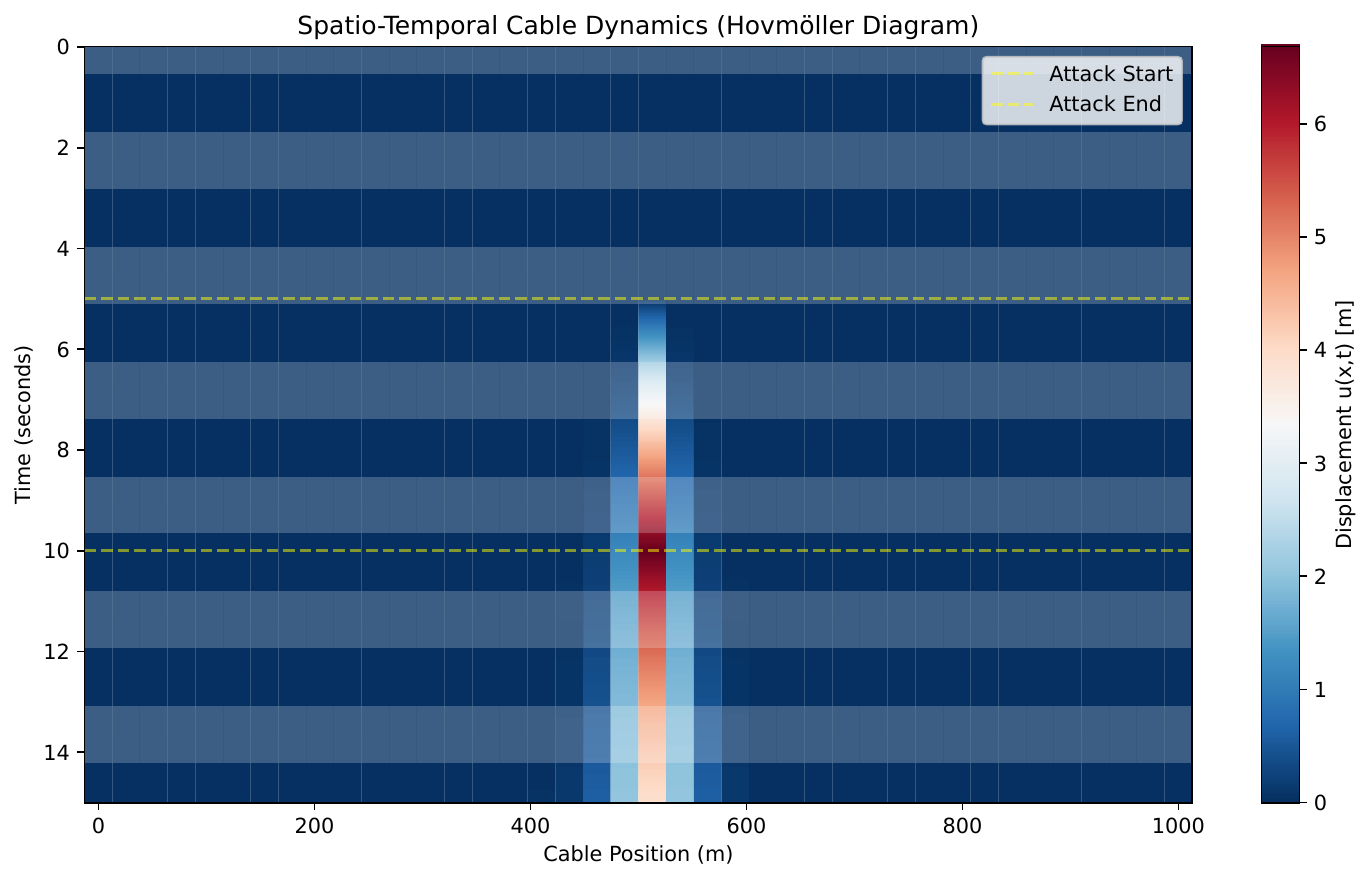}
    \caption{Spatio-temporal cable dynamics. The color-map depicts the 
    emergence of the localized tampering event and the subsequent 
    wave-based relaxation phase.}
    \label{fig:dynamics}
\end{figure}

\noindent {\bf Simulation Results and Analysis.}
The proposed framework is evaluated through a simulated $5\,$s tampering event. Specifically, we analyze the resulting physical and statistical behaviors.
Regarding the physical dynamics,
the spatio-temporal evolution of the cable displacement is illustrated 
in the Hovmöller diagram in Fig.~\ref{fig:dynamics}, which is a space–time plot showing the
evolution of the displacement field along the cable over time.
The figure reveals a clear 
localized deformation centered at $x=500\,$m during the attack window. 
Upon removal of the external force at $t=10\,$s, the diagram captures 
the characteristic wave reflections and damped oscillations. 
These transient oscillatory effects result from the energy stored in the 
deformed tensioned cable, which propagates toward the fixed boundaries 
before being dissipated by hydrodynamic drag.

\begin{figure}[htpb]
    \centering
    \includegraphics[width=0.98\columnwidth]{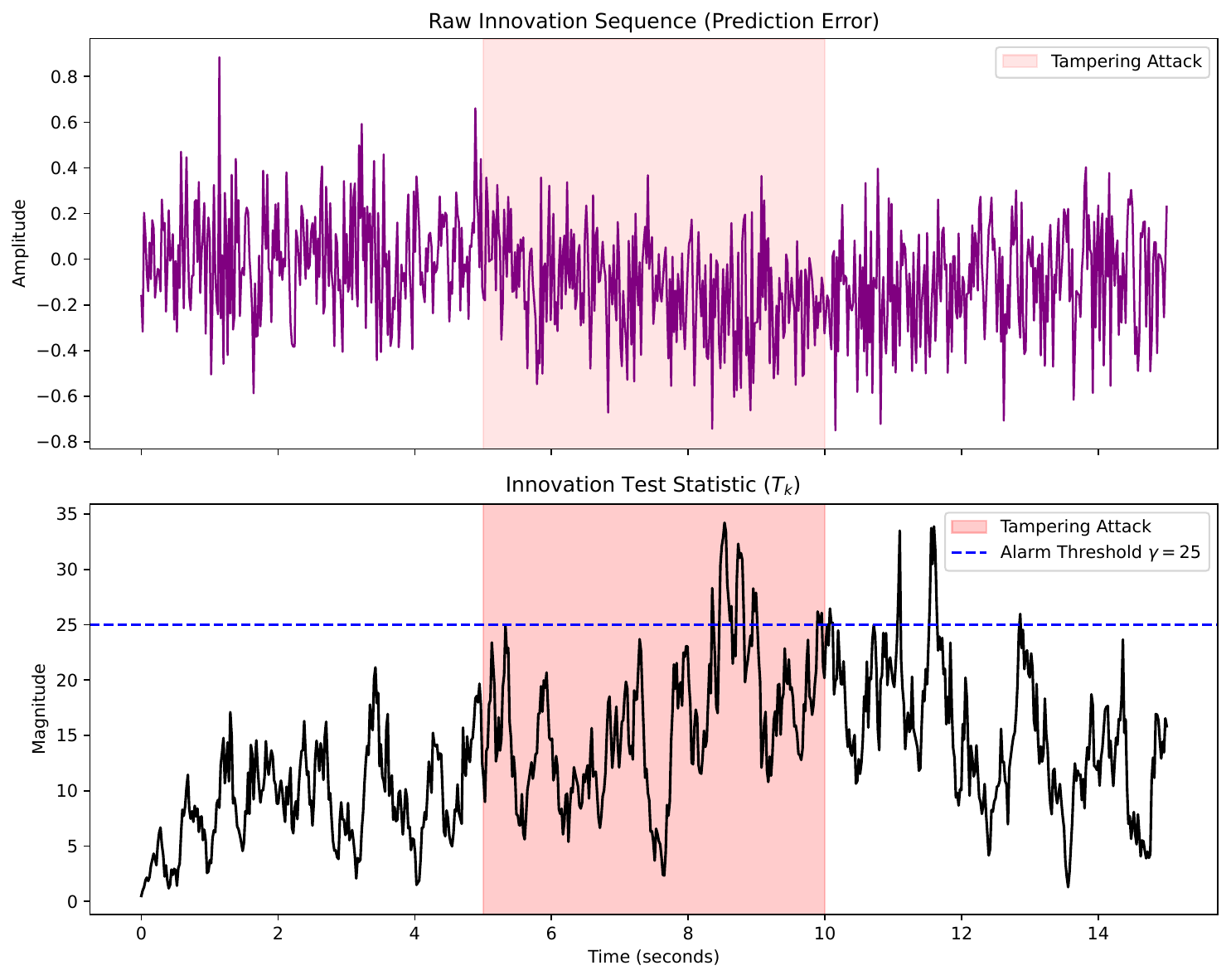}
    \caption{Innovation-based detection results. (Top) Central channel 
    innovation sequence. (Bottom) $\chi^2$ test statistic showing 
    robust detection of the tampering event.}
    \label{fig:stats}
\end{figure}

We next consider detection performance; the statistical response is summarized in Fig.~\ref{fig:stats}. 
The Kalman filter, which is initialized with maximum uncertainty, undergoes 
 a brief startup transient before settling into a zero-mean innovation 
state under nominal ocean noise. During the tampering window, the 
innovation test statistic $T_k$ exhibits a clear exceedance of the 
threshold $\gamma=25$, demonstrating high sensitivity to the unmodeled 
external force. The secondary peaks observed post-attack ($t > 10\,$s) 
align with the physical oscillations identified in the Hovmöller diagram, 
correctly signaling that the system state has not yet returned to 
the stochastic equilibrium.

%
Notice that the robustness of the detection scheme relies on the tuning of the process and measurement covariances. Increasing the expected process noise intensity $Q$ reduces the filter's confidence in the dynamic model, which slightly decreases the sensitivity to minor tampering but strongly suppresses false alarms. Conversely, finer spatial discretization $N$ improves the physical localization of the event at the cost of higher computational burden during the matrix inversion steps. Overall, the proposed approach exhibits stable behavior in this representative simulated setting under these physically grounded parameters. The present validation is intentionally limited to simulation and does not yet compare against alternative data-driven or threshold-only detectors. Such comparisons, together with experimental calibration on real distributed-sensing data, are important next steps for quantifying field performance.
\section{Conclusion}

In this paper, we present and assess in simulation a framework for the real-time detection of anomalous mechanical disturbances in submarine cables using distributed deformation measurements. By utilizing a physical wave-equation model and a Kalman filtering architecture, we successfully separated stochastic environmental forcing from systematic external perturbations acting on the cable. The proposed formulation provides an interpretable innovation signal and an explicit statistical decision rule, while relying on modeling assumptions whose range of validity must be checked for each deployment.

Future work directions will focus on the classification of detected anomalies using machine learning techniques to differentiate between various disturbance sources, such as anchor strikes, seabed interactions, or nearby vessel activity. In addition, the proposed monitoring framework could be implemented
using distributed fiber sensing technologies, such as Distributed
Acoustic Sensing (DAS), which enable spatially distributed strain or
vibration measurements along long cable structures. Extending the physical model to incorporate non-linear damping and heterogeneous seabed coupling conditions will further enhance the applicability of the approach to realistic maritime environments. Future work will address experimental validation and covariance \mbox{calibration} from nominal data, along with benchmark comparisons against alternative anomaly detectors.

\bibliographystyle{IEEEtran}
\bibliography{references}

\end{document}